\documentclass{article}
\usepackage{amsmath}
\usepackage[psamsfonts]{amssymb}
\usepackage{cmmib57}
\usepackage{diagrams}
\newcommand{\bPf}{\par\vspace*{-4pt}\indent{\sc Proof.}\enskip}
\newcommand{\ePf}{\medskip}
\def\QED{\hskip0.1em\hfill\null\ \null\nobreak\hfill\kern3pt\vbox{\hrule\hbox
   {\vrule\kern1pt\vbox{\kern1.7pt\hbox{$\scriptscriptstyle{QED}$}
    \kern0.2pt}\kern1pt\vrule}\hrule}}

\def\END{\hskip0.1em\hfill\null\ \null\nobreak\hfill\kern3pt\vbox{\hrule\hbox
   {\vrule\kern1pt\vbox{\kern1.7pt\hbox{$\,\,\,\vspace{5pt}$}
    \kern0.2pt}\kern1pt\vrule}\hrule}}
\newtheorem{theorem}{Theorem}
\newtheorem{lemma}{Lemma}
\newtheorem{corollary}{Corollary}
\newtheorem{proposition}{Proposition}
\newtheorem{remark}{Remark}
\newtheorem{definition}{Definition}
\newtheorem{example}{Example}
\newcommand{\bCd}{\bEq\begin{CD}}
\newcommand{\eCd}{\end{CD}\eEq}
\newcommand{\bcd}{\beq\begin{CD}}
\newcommand{\ecd}{\end{CD}\eeq}
\newcommand{\ben}{\begin{enumerate}}
\newcommand{\een}{\end{enumerate}}
\newcommand{\bEq}{\begin{eqnarray}}
\newcommand{\eEq}{\end{eqnarray}}
\newcommand{\beq}{\begin{eqnarray*}}
\newcommand{\eeq}{\end{eqnarray*}}
\newcommand{\bDf}{\begin{definition}\em}
\newcommand{\eDf}{\end{definition}}
\newcommand{\bLm}{\begin{lemma}}
\newcommand{\eLm}{\end{lemma}}
\newcommand{\bPr}{\begin{proposition}}
\newcommand{\ePr}{\end{proposition}}
\newcommand{\bTh}{\begin{theorem}}
\newcommand{\eTh}{\end{theorem}}
\newcommand{\bCr}{\begin{corollary}}
\newcommand{\eCr}{\end{corollary}}
\newcommand{\bRm}{\begin{remark}\em}
\newcommand{\eRm}{\end{remark}}
\newcommand{\bEx}{\begin{example}\em}
\newcommand{\eEx}{\end{example}}

\newcommand{\ie}{{\em i.e$.$} }
\newcommand{\eg}{{\em e.g$.$} }

\newcommand{\mto}{\mapsto}

\newcommand{\der}{\partial}

\DeclareMathOperator{\byd}{{\raisebox{.1ex}{:}{=}}}

\newcommand{\ucar}[1]{\underset{#1}{\times}}

\newcommand{\cC}{\mathcal{C}}
\newcommand{\cD}{\mathcal{D}}

\newcommand{\bB}{\boldsymbol{B}}

\newcommand{\bK}{\boldsymbol{K}}

\newcommand{\bU}{\boldsymbol{U}}

\newcommand{\bX}{\boldsymbol{X}}

\newcommand{\bZ}{\boldsymbol{Z}}

\newcommand{\sub}{\subset}

\newcommand{\wed}{\wedge}
\newcommand{\com}{\!\circ\!}

\newcommand{\alp}{\alpha}
\newcommand{\bet}{\beta}

\newcommand{\tht}{\theta}

\newcommand{\sig}{\sigma}

\newcommand{\ome}{\omega}

\newcommand{\Ome}{\Omega}

\newcommand{\Tht}{\Theta}

\title{\large{{\bf Nonlinear $2+1$--Dimensional Field Equations from Incomplete Lie Algebra
Structures}}}
\bigskip
\author{\large{Marcella Palese}\thanks{Supported by GNFM of INdAM, MURST and University of
Torino.}\, and \large{Ekkehart Winterroth}\thanks{
On leave of absence from Department of Mathematics,
University of Erlangen--N\"urnberg, Bismarkstrasse 1 1/2, 
Erlangen, Germany. 
}
\\{\footnotesize Department of Mathematics,
University of Torino}
\\{\footnotesize Via C. Alberto 10, 10123 Torino, Italy}
\\{\footnotesize e--mails: {\sc marcella.palese@unito.it, 
ekkehart.winterroth@edu.dm.unito.it}}}
\date{}
\begin{document}

\maketitle

\begin{abstract}
We show that the nonlinear $2+1$--dimensional Three--Wave Resonant Interaction
equations, describing several important physical phenomena,
can be generated starting from incomplete Lie algebras in the framework of multidimensional
prolongation structures. We make use of an {\em ansatz} involving the structure
equations of a principal prolongation connection induced by an admissible B\"acklund map.

\medskip

{\footnotesize \noindent \small{{\bf Key words}: exterior differential systems, integrable field
equations, prolongation Lie algebras, B\"acklund transformations, principal connections, fibered
manifolds.

}}

\end{abstract}


\small

\section{Introduction}

As it is well known, completely integrable nonlinear field equations admit
Lax pairs, (multi--)soliton solutions and an infinite set of conservation laws.
It was pointed out that all these properties can be related with the
existence of specific kinds of B\"acklund transformations (see {\em e.g.} \cite{AbSe81,AI79}).

In this context the prolongation structures method, which can
be interpreted as the construction of an Ehresmann connection \cite{Eh51,PRS79}, plays a relevant
role (see \eg \cite{Es82,Es89,Es90,He76,He76b,Wa75}). One of the most interesting features of the
arising algebraic structures is that they are  homomorphic to infinite--dimensional {\em loop}
algebras  (see {\em e.g.}
\cite{ALLPS94,ALLPS95,PALLS96,Pa93} and references quoted therein).  The inverse
prolongation structures procedure can be thought as the search of the B\"acklund map which
generates a given (incomplete) Lie algebra structure \cite{Es82,Es89,Es90,He76,PRS79,Wa75}. 
This inverse procedure, based on the Cartan method of moving frames, was first outlined by
Estabrook \cite{Es82,Es89,Es90} who showed how exterior differential systems can be obtained from
incomplete Lie algebra structures. Further results concerning the (1+1)--dimensional 
case were exploited \eg in \cite{ALLPS94,ALLPS95,Es82,Es89,Es90,Ho86,LeSo89,PALLS96,Pa93}.
Starting from a given incomplete Lie algebra structure, it is 
then possible to generate the field equations whose prolongation structure 
is the given algebra\footnote{As well as the associated linear spectral 
problem useful for the integration {\em via} the Inverse Spectral Transform (IST) method. For a
review of the IST method see \eg \cite{FoGe94} and references quoted therein.}. In fact, the
extension of both the direct and inverse prolongation  structure procedure to the
multi--dimensional case is not trivial and presents several technical difficulties (see \eg
\cite{Pa93,PLS00}).

This note is aimed to give a contribution in the geometrical characterization of {\em integrability
properties} of nonlinear field equations in the framework of the inverse procedure in $2+1$
dimensions. It is in fact well known that the study of higher dimensional systems is a central
theme in the theory of integrable systems, nevertheless the investigation was always made in the
direction of extension of (1+1)--dimensional systems to (2+1)--dimensional ones just {\em via} the
extension either of the Lax pair (see \eg
\cite{AbHa75}) or of the so--called  prolongation forms (see \eg \cite{Mo76,To85}), as
well as of the moving frames setting \cite{LMVN98}. In this paper the more general approach of
{\em generating new (2+1)--dimensional integrable systems from a given abstract algebraic structure
via the extension of a B\"acklund map} is tackled.

We are, in general, interested in the case when incomplete Lie algebras 
arise as necessary conditions for integrability of a 
(formal) connection induced by a B\"acklund map \cite{PRS79}, and the 
choice of a realization of such algebras (particularly in 
loop--algebras form) 
gives us the solution of the so--called B\"acklund problem. 
Conversely, if the algebraic structure is known and the ` new' variable (also called
pseudopotentials \cite{Wa75}) dependence of the B\"acklund map is fixed, the integrability 
condition for the formal connection induced by a B\"acklund map 
provides the whole family of exterior differential systems which 
admits the given B\"acklund map.

We shall make
use of an {\em ansatz} which is based on the fact that, from a geometric point of 
view, the connection forms (induced by an admissible B\"acklund map) play the relevant role. This
approach is then slightly different from the one 
formulated in the
$2+1$--dimensional direct prolongation procedure by Morris and Tondo 
\cite{Mo76,To85} and it should be seen within the geometric approach in the
framework of jet bundles and connections provided by Pirani {\em et al.} 
(see \eg \cite{PRS79}). The {\em ansatz} is in fact a slightly modified version 
of the structure equations of the connection, which will be the starting point for the 
application of the inverse method. According to this, in Section \ref{2} we recall 
how a connection can be induced by a B\"acklund map in the jet bundles framework and provide a
characterization of completely integrable systems in terms of B\"acklund structures. 

In Section \ref{3}, we give an example of physical application of 
such a generalized {\em ansatz} providing some nonlinear 
$2+1$--dimensional field equations describing various physical 
phenomena \cite{Mo76,To85,Ka80,Be62,Be67,AmBe00}. We show that the 
$2+1$--dimensional Three Wave Resonant Interaction (3WRI) equations, together 
with a condition on the group velocities which is related with the 
resonance condition, can be generated from an 
incomplete Lie algebra, as the integrability condition for a special 
class (which we shall call {\em admissible}) of B\"acklund connections.
Our {\em ansazt} enables us on the other hand to generate a whole family of nonlinear field
equations related with the $3$WRI equations by transformations of coordinates and fields of the
Miura type. All the family is in fact `contained' in the exterior differential system admitted by
a postulated (admissible) B\"acklund map as the integrability condition of the corresponding induced
B\"acklund connection.

\section{B\"acklund transformations and induced Connections}\label{2}

In the following we shall shortly recall few basics concepts and set the notation.
We shall assume the reader is familiar with the basic notions from 
the theory of bundles, jet prolongations, principal bundles 
and connections (for references and details see \eg \cite{Ko72,KoNo63,KMS93}).

\medskip

Let $\pi: \bU\to \bX$, $\tau: \bZ\to \bX$,
be two (vector) bundles 
with local fibered coordinates $(x^{\alp},u^{A})$ and $(x^{\alp},z^{i})$,
respectively, where $\alp=1,\ldots,m=\textstyle{dim}\bX$, $A=1,\ldots,n=\textstyle{dim}\bU -
\textstyle{dim}\bX$, $i=1,\ldots,N=\textstyle{dim}\bZ-\textstyle{dim}\bX$. We shall assume that
sections of the bundles $\bU=(\bU,\bX,\pi)$ and $\bZ=(\bZ,\bX,\tau)$ represent physical fields
sastisfying a given system of (nonlinear) field equations. 

\medskip

\medskip

A system of nonlinear field equations of order $k$ on $\bU$ is geometrically described as an
exterior differential system $\nu$ on $J^{k}\bU$
\cite{Es82,Es89,Es90,He76,He76b,KMS93,Sau89,Wa75}. The solutions of the field equations are (local) 
sections
of $\bU\to\bX$ -- \ie (local) mappings $\sig:\bX\to\bU$ such that $\sig\com\pi=
\textstyle{id}_{\bX}$ -- such that $(j^{k}\sig)^{*}\nu=0$. They define a submanifold in $J^{k}\bU$. 
We shall also denote by
$J^{\infty}\nu$ ({\em resp.} $j^{\infty}\sig$) the infinite order jet prolongation of $\nu$ 
({\em resp.} $\sig$). 

\subsection{Admissible B\"acklund transformations}

We recall that the group of {\em contact transformations} of a bundle is the group of its
infinitesimal fibered automorphisms.
A contact transformation preserves then, by definition, the fibering (see \eg \cite{KMS93,Sau89}).
Let then $\bB$ be the infinite--order contact transformations group on $J^{\infty}\bU$ \cite{AI79}.

\medskip

In the following we recall some basic definitions (see \eg \cite{AI79}) and stress some important
properties which will be used later.

\bDf\label{Backlund}
The group of {\em B\"acklund transformations for the system $\nu$} is the closed subgroup $\bK$ of
$\bB$ which leaves invariant $J^{\infty}\nu$. 
\END\eDf

\bDf\label{admissible}
The group of {\em admissible B\"acklund transformations for the system $\nu$} is the closed subgroup
$\Tilde{\bK}$ of $\bB$ which leaves invariant solution submanifolds of $J^{\infty}\nu$ . 
\END\eDf

\bRm The contact transformations group $\bB$ acts freely on
the infinite dimensional submanifold of $J^{\infty}\bU$ defined by $J^{\infty}\nu$. 
The group $\Tilde{\bK}$ of admissible B\"acklund transformations 
is then the compact subgroup of $\bB$ which preserves contact elements along the solutions of the
given (system of) nonlinear field equations.
Thus $\Tilde{\bK}$ can be seen as the isotropy subgroup of $\bB$.
\END\eRm

Let $\pi: \bU\to\bX$, $\tau:\bZ\to \bX$, be vector bundles as the above and 
$\pi^{1}: {J^{1}\bU} \to {\bX}$, $\tau^{1}: {J^{1}\bZ} \to {\bX}$, the first order jet
prolongations bundles\footnote{A B\"acklund transformation can be analogously defined at any jet
order \cite{PRS79}, but here we will consider only the jet order which is involved with the
physical application we are concerned with.}, with local fibered coordinates
$(x^{\alp},u^A,u^{A}_{\alp})$,
$(x^{\alp},z^i,z^{i}_{\alp})$, respectively. Furthermore, let
$(\partial_{\beta}$, $\partial_{A}$, $\partial_{A}^{\beta})$,
$(\partial_{\beta},\partial_{i}$, $\partial_{i}^{\beta})$ and $(dx^{\beta}$,
$du^{A}$, $du^{A}_{\beta})$, $(dx^{\beta},dz^{i}$, $dz^{i}_{\beta})$ be local
bases of tangent vector fields and $1$--forms on $J^{1}\bU$ and $J^{1}\bZ$, respectively. 
In the sequel we will be concerned with the pull--backs $\pi^*(\bZ)\simeq \bZ\ucar{\bX}\bU$, 
$\tau^*(\bU)\simeq \bU\ucar{\bX}\bZ$, 
$\eta^*(J^{1}\bU)\simeq J^{1}\bU\ucar{\bX}\bZ$, 
where $\eta\byd\tau^*(\pi)$.

\medskip

\bDf
Following \cite{PRS79}, we define a B\"acklund map to be the fibered morphism over $\bZ$:
\bEq
\phi: J^{1}\bU\ucar{\bX}\bZ\to J^{1}\bZ: (x^{\alp},u^A,u^{A}_{\alp};z^{i})\mto
(x^{\alp},z^{i},z^{i}_{\alp})\,,
\eEq
with $z^{i}_{\alp}=\phi^{i}_{\alp}(x^{\bet},u^A,u^{A}_{\bet};z^{j})$.
\END
\eDf

\bDf
The fibered morphism $\phi$ is said to be an {\em admissible} B\"acklund
transformation for the differential system $\nu$ if
$z^{i}_{\alpha}=\phi^{i}_{\alpha}(x^{\beta},u^{A},u^{A}_{\beta};z^{j})$, 
$\phi^{i}_{\alpha}=\cD_{\alpha}\phi^{i}$ and the integrability
conditions 
\bEq\label{integrabilityI}
\cD_{\alpha}\phi^{j}_{\ome} =
\cD_{\ome}\phi^{j}_{\alpha} \label{curvature}\,,
\eEq 
(with $\cD_{\alpha} = \partial_{\alpha} + u^{A}_{\alpha}\partial_{A}+
u^{A}_{\alpha\beta}\partial_{A}^{\beta}+
\phi^{i}_{\alpha}\partial_{i}\label{invariant}$) coincide with the
exterior differential system $\nu$ \cite{AI79,PRS79}. 
\END\eDf

\bRm 
By pull--back of the contact structure on $J^1\bZ$, the B\"acklund morphism induces in a
natural way an horizontal (with respect to $\pi_0^{1 \, *}$) distribution on the bundle
$(J^{1}{\bU}\times_{\bX}{\bZ},$ $J^{1}{\bU},$ $\pi^{1*}_{0}(\eta))$.
The local expression of generators of such a horizontal distribution is given by the following
connection forms 
\bEq\label{horizontal} \Tht^{i}=
dz^{i}-\phi^{i}_{\beta}(x^{\alpha},u^A,u^{A}_{\alpha};z^{j})dx^{\beta}\,. \eEq
\END\eRm

\bDf
The horizontal distribution locally defined by Eq. \eqref{horizontal} is called the {\em induced
B\"acklund connection}.
\END
\eDf

\bDf\label{integrability} 
The system $\nu$ is said to be {\em completely integrable} if there exists a normal subgroup
$\bK_{0}\sub (\Tilde{\bK} \cap \bK) \sub \bB$ leaving invariant (the infinite order
prolongation of) $\nu$ and its solutions.
\END\eDf

Let $\chi_{k}=\chi^{i}_{k}\der_{i}$ be generators of the Lie algebra
$\mathfrak{k_{0}}$ of the Lie subgroup $\bK_{0}$, which satisfy the commutation relations 
\bEq 
\left [\chi_{l},\chi_{m}\right ] = - \, {\cal C}^{k}_{lm}\chi_{k} \,.
\eEq
where ${\cC}^{n}_{lm}$ are the structure constants of the Lie algebra
$\mathfrak{k_{0}}$.

\bTh
The following statements are equivalent.
\begin{enumerate}
\item $\phi$ is an admissible B\"acklund transformation for the
differential system $\nu$. 
\item The horizontal
distribution \eqref{horizontal} is
$\bK_{0}$--invariant. 
\end{enumerate}
\eTh

\bPf 
If the horizontal
distribution \eqref{horizontal} is
$\bK_{0}$--invariant, we have (see \eg \cite{KMS93}):
\bEq \label{Lie}
\phi^{i}_{\beta}(x^{\alpha}, u^{A}, u^{A}_{\alpha}; z^{j}) =
\omega^{k}_{\beta}(x^{\alpha}, u^{A}, u^{A}_{\alpha})\chi^{i}_{k}(z^{j}) \,.
\eEq 
Then the integrability conditions \eqref{integrabilityI} for the B\"acklund map $\phi$ are
equivalent to the integrable exterior differential system:
\bEq\label{system}
\nu\byd\Ome^{k}_{\alp\bet} =0\,,
\eEq
where 
\beq
\Ome^{k}_{\alp\bet}=\cD_{\alpha}\ome^{k}_{\bet}-\cD_{\alpha}\ome^{k}_{\alp}+
{\cal C}^{k}_{lm}\ome^{l}_{\alp}\ome^{m}_{\bet}\,.
\eeq
In fact, taking Eq.s \eqref{horizontal} into account, from the equivariance condition, the structure
equations for the connection induced by the admissible B\"acklund map are (up to pull--backs):
\bEq
d\Tht^{i} & = & (d\ome^{k}+\frac{1}{2}{\cC}^{k}_{lm}\ome^{l}\wed\ome^{m})\chi^{i}_{k}
\qquad (\textstyle{mod}\,\Tht^{j})\label{boh1}\\
& = & \frac{1}{2}\Ome^{k}_{\alp\bet}\chi^{i}_{k}\,dx^{\alp}\wed dx^{\bet}\label{boh2} \,,
\eEq
where $\ome^{k}=\ome^{k}_{\alp}dx^{\alp}$ are $1$--forms on $J^{1}\bU$, horizontal with respect to
$\pi^{1}$. The integrability condition $d\Tht^{i}=0$ is then equivalent to
$\Ome^{k}_{\alp\bet}\chi^{i}_{k}=0$, \ie to Eq. \eqref{system}.

Conversely, if the B\"acklund map is admissible for the system $\nu$, then the group $\bK_{0}$
leaves invariant the solution (integral maximal) submanifolds of $\nu$, then the horizontal
distribution induced by the B\"acklund map is $\bK_{0}$--invariant.
\QED
\ePf

\bCr\label{cacca}
A nonlinear exterior system $\nu$ is completely integrable if and only if there exists an
admissible B\"acklund transformation $\phi$ for the differential system $\nu$.
\eCr

\bPf
It follows from Definition \ref{integrability} and the above Theorem.
\QED\ePf

\section{The $(2+1)$-3WRI from incomplete Lie algebras}\label{3}

From a geometric point of 
view, the connection forms induced by an admissible B\"acklund map play a relevant role. Once that
the dependence of the connection forms on the $z^{i}$ variables is known\footnote{This is the case
when any abstract incomplete Lie algebra structure is given.}, the integrability condition (the
{\ie \em zero curvature} condition) for such a connection provides,
{\em via} the inverse procedure and thanks to Corollary \ref{cacca}, whole families of {\em
integrable} differential systems
\cite{ALLPS94,ALLPS95,Es82,Es89,Es90,Ho86,LeSo89,PALLS96,Pa93}.

This suggests the introduction of an {\em ansatz} which is aimed to
generalize the geometric approach in the framework of jet bundles provided by Pirani {\em et al.}
\cite{PRS79} and Hoenselaers \cite{Ho86}. The {\em ansatz} is in fact a slightly modified
version  of the structure equations, which will be the starting point for the 
application of the inverse method.

\medskip

Let us then consider the following {\em ansatz}:
\bEq\label{ansatz}
d\Ome^{k}=0\,,  \qquad\textstyle{with} \qquad \Ome^{k}=\tht\wed\Tht^{k}\,,
\eEq
where $\tht$ is a closed $1$--form on $J^{1}\bU$ horizontal over $\bX$, 
and $\Tht^{k} =
dz^{k}+\chi^{k}_{j}(z^{m})\ome^{j}$
are {\em admissible B\"acklund connection forms} on
$J^{1}\bU\ucar{\bX}\bZ$, with $k=1,\ldots,N\,= \,\textstyle{dim}\,\mathfrak{b}$, $j=1,\ldots, M\,=
\,\textstyle{dim}\,\mathfrak{k_{0}}$\footnote{The geometrical interpretation of this 
{\em ansatz}, will be investigated in a separate paper
\cite{PaWi01}.}. 

Assume now that the dependence on the $z^{i}$ variables is provided by the
following incomplete\footnote{It is incomplete in the sense that not all of the commutators are
known, then the algebra is not closed as a Lie algebra structure.} Lie algebra structure

\beq 
\begin{array}{lcllcllcl} \left[\chi_{1},\chi_{5}\right] & = & 0 &
\left[\chi_{2},\chi_{4}\right] & = & 0 & \left[\chi_{1},\chi_{6}\right] & = & 0 \\
\left[\chi_{3},\chi
_{4}\right] & = & 0 & \left[\chi_{2},\chi_{6}\right] & = & 0 &
\left[\chi_{3},\chi_{5}\right] & = & 0 \\ \left[\chi_{1},\chi_{7}\right] & = & 0 &
\left[\chi_{4},\chi_{7}\right] & = & 0 & \left[\chi_{2},\chi_{8}\right] & = & 0 \\
\left[\chi_{5},\chi_{8}\right] & = & 0 & \left[\chi_{7},\chi_{8}\right] & = & 0 & &   &   \\
\left[\chi_{1},\chi_{2}\right] & = & a\chi_{6} & \left[\chi_{1},\chi_{3}\right] & = & b\chi_{5} &
\left[\chi_{2},\chi_{3}\right] & = & c\chi_{4} \\ \left[\chi_{4},\chi_{5}\right] & = & -a\chi_{3} &
\left[\chi_{4},\chi_{6}\right] & = & -b\chi_{2}& \left[\chi_{5},\chi_{6}\right] & = & -c\chi_{1} \\
\left[\chi_{3},\chi_{7}\right] & = & \left[\chi_{3},\chi_{8}\right],&
\left[\chi_{6},\chi_{7}\right] & = & \left[\chi_{6},\chi_{8}\right],& &   &   \\ 
\end{array}
\eeq 
where $a = \frac{i}{\lambda_{12}}$, $b = \frac{i}{\lambda_{13}}$, $c =
\frac{i}{\lambda_{23}}$, with $\lambda_{12}$, $\lambda_{23}$, $\lambda_{13}$
different from zero.

\medskip

Thus, by requiring the structure equations \eqref{boh2} hold true, from the {\em ansatz}
\eqref{ansatz}, we obtain the following {\em integrable} exterior differential system:

\bEq 
\theta \wedge d\ome^{7} =  0 \,,\quad \theta \wedge d\ome^{8} = 0 \,,
\label{gamma7} \eEq \bEq \theta \wedge (d\ome^{1}+ c\ome^{5}\wedge\ome^{6})
= 0 \,,\quad \theta \wedge (d\ome^{2} + b\ome^{4}\wedge\ome^{6}) = 0 \,,
\label{dgamma1} \eEq \bEq \theta \wedge (d\ome^{3} +
a\ome^{4}\wedge\ome^{5}) = 0 \,, \quad \theta \wedge (d\ome^{4} -
c\ome^{2}\wedge\ome^{3}) = 0 \,, \eEq \bEq \theta \wedge (d\ome^{5} -
b\ome^{1}\wedge\ome^{3}) = 0 \,, \quad \theta \wedge (d\ome^{6} -
a\ome^{1}\wedge\ome^{2}) = 0Ê\,, \label{dgamma6} \eEq \bEq \theta \wedge
\ome^{3}\wedge (\ome^{7}-\ome^{8}) = 0 \, \quad \theta \wedge
\ome^{6}\wedge(\ome^{7}-\ome^{8}) = 0 \,,\label{b} \eEq \beq \theta \wedge
\ome^{1} \wedge \ome^{4} = \theta \wedge \ome^{1} \wedge \ome^{8} =
\theta \wedge \ome^{2} \wedge \ome^{5} = \eeq \bEq \theta \wedge \ome^{2}
\wedge \ome^{7} =Ê \theta \wedge \ome^{3} \wedge \ome^{6} = \theta \wedge
\ome^{4} \wedge \ome^{8} = \theta \wedge \ome^{5} \wedge \ome^{7} =
0\label{c}\,. 
\eEq 

The choice of fibrations of $\bU$ over a basis
manifold provides a whole family of nonlinear fields equations related by Miura's transformations
(see \eg \cite{ALLPS94,ALLPS95,LeSo89,PALLS96,Pa93}). In the sequel we will
find out the 3WRI equations making a specific choice. 

\medskip

Along a section of
$\bZ\to\bX$ the closed form
$\tht$ can be chosen as 
\bEq \theta = m_{1}dx + m_{2}dy +
m_{3}dt\label{sold}\,,
\eEq 
where $m_{i}$, ($i=1,2,3$) are some constants\footnote{In this way we obtain a case which turns
out to be degenerate in the standard AKNS formulation (see \eg \cite{AbHa75} for further
details).}.  Then \eqref{gamma7} imply
\bEq
\ome^{7} & = & n_{1}dx + n_{2}dy + n_{3}dt\,, \\ \ome^{8} & = & p_{1}dx +
p_{2}dy + p_{3}dt \,, 
\eEq 
with $n_{i},p_{i}$, ($i=1,2,3$)
constants. Moreover, from \eqref{c}, we have that \bEq \ome^{1} & = & u
\ome^{8} \,, \label{miura1}\qquad \ome^{2} = v \ome^{7} \,,\\ \ome^{3} &
= & r (\ome^{7}-\ome^{8}) \,,\qquad \ome^{4} = z \ome^{8} \,,\\
\ome^{5} & = & w \ome^{7} \,,\qquad \ome^{6} = s (\ome^{7}-\ome^{8})
\,.\label{miura2} 
\eEq

Let us now consider $r$, $s$, $u$, $v$, $z$ and $w$ as complex
functions of the independent variables (this is equivalent to a choice of a fibration of $\bU$ over
the manifold $\bX$) as follows:
\bEq 
u & \byd & u_{1}(x, y, t) \,,
\quad v \byd u_{2}(x, y, t) \,, \quad r \byd u_{3}(x, y, t) \,, \label{sub1}\\ z & \byd &
u^{*}_{1}(x, y, t) \,, \quad w \byd u^{*}_{2}(x, y, t) \,, \quad s \byd u^{*}_{3}(x, y,
t) \,. 
\eEq
Furthermore, for notational convenience, we put:
\bEq  
a_{1} & = &  
\frac{m_{3}p_{2} - m_{2}p_{3}}{m_{2}p_{1}
- m_{1}p_{2}}\,, \quad b_{1} = \frac{m_{1}p_{3} - m_{3}p_{1}}{m_{2}p_{1} -
m_{1}p_{2}} \,, \label{syst1}\\ a_{2} & = & \frac{m_{3}n_{2} - m_{2}n_{3}}{m_{2}n_{1}
- m_{1}n_{2}} \,, \quad b_{2} = \frac{m_{1}n_{3} - m_{3}n_{1}}{m_{2}n_{1} -
m_{1}n_{2}} \,, \\ a_{3} & = & \frac{m_{3}q_{2} - m_{2}q_{3}}{m_{2}q_{1} -
m_{1}q_{2}} \,, \quad b_{3} = \frac{m_{1}q_{3} - m_{3}q_{1}}{m_{2}q_{1} -
m_{1}q_{2}} \,, 
\eEq 
and
\bEq 
\lambda_{23} & = & n_{3} + \frac{m_{3}p_{2} -
m_{2}p_{3}}{m_{2}p_{1} - m_{1}p_{2}}n_{1} + \frac{m_{1}p_{3} -
m_{3}p_{1}}{m_{2}p_{1} - m_{1}p_{2}}n_{2}\,,\\ 
\lambda_{13}  & = &  - p_{3} -
\frac{m_{3}n_{2} - m_{2}n_{3}}{m_{2}n_{1} - m_{1}n_{2}}p_{1} - \frac{m_{1}n_{3} -
m_{3}n_{1}}{m_{2}n_{1} - m_{1}n_{2}}p_{2}  \,,\\ 
\lambda_{12}  & = & - n_{3} -
\frac{m_{3}q_{2} - m_{2}q_{3}}{m_{2}q_{1} - m_{1}q_{2}}n_{1} - \frac{m_{1}q_{3} -
m_{3}q_{1}}{m_{2}q_{1} - m_{1}q_{2}}n_{2}\,, \label{syst9}
\eEq 
where $\lambda_{12}= a_{1}b_{2} - a_{2}b_{1}$, $\lambda_{23}=a_{2}b_{3} -
a_{3}b_{2}$, $\lambda_{13} = a_{1}b_{3} - a_{3}b_{1}$, and $q_{i} = n_{i} -
p_{i}$, $i=1,2,3$.

Then Eqs. \eqref{dgamma1}--\eqref{dgamma6} with \eqref{sold}--\eqref{miura2} provide
the following system of integrable NFEs 
\bEq u_{1t} + a_{1}u_{1x} + b_{1}u_{1y} & = &
iu_{2}^{*}u_{3}^{*} \,, \label{Eq1}\\ u_{2t} + a_{2}u_{2x} + b_{2}u_{2y} & = &
iu_{1}^{*}u_{3}^{*} \,, \\ u_{3t} + a_{3}u_{3x} + b_{3}u_{3y} & = &
iu_{1}^{*}u_{2}^{*} \,,\label{Eq3} 
\eEq 
together with their complex conjugates. 

\bRm The compatibility condition for the system of Eqs. \eqref{syst1}--\eqref{syst9} is 
\bEq
\lambda_{12} + \lambda_{23} = \lambda_{13}\,.\END
\label{resonance} 
\eEq \eRm

\bRm Eqs. \eqref{Eq1}--\eqref{Eq3}, together with their complex conjugate, are
just the equations describing  the resonant interaction of the three waves
envelope in $(2+1)$ dimensions \cite{Ka80,Mo76,To85}. Equation \eqref{resonance}
is related to the resonance condition for the wave envelope (see \eg \cite{Ka80}). \END\eRm

\subsection{Conclusions}

We provided a characterization of completely
integrable nonlinear field equations in terms of algebraic properties of
associated B\"acklund structures by investigating the relation between
B\"acklund transformations and connections theory. As an example of
application, by resorting to the structure equations
induced by an integrable admissible B\"acklund map, we have shown how nonlinear $(2+1)$--dimensional
field equations can be generated starting from
incomplete Lie algebras.


\end{document}